# A General-Purpose Machine Learning Framework for Predicting Properties of Inorganic Materials

Logan Ward[1], Ankit Agrawal[2], Alok Choudhary[2], Christopher Wolverton[1]

[1] Department of Materials Science and Engineering, Northwestern University, Evanston, IL, USA

[2] Department of Electrical Engineering and Computer Science, Northwestern University, Evanston, IL, USA

## Abstract

A very active area of materials research is to devise methods that use machine learning to automatically extract predictive models from existing materials data. While prior examples have demonstrated successful models for some applications, many more applications exist where machine learning can make a strong impact. To enable faster development of machine-learning-based models for such applications, we have created a framework capable of being applied to a broad range of materials data. Our method works by using a chemically diverse list of attributes, which we demonstrate are suitable for describing a wide variety of properties, and a novel method for partitioning the data set into groups of similar materials in order to boost the predictive accuracy. In this manuscript, we demonstrate how this new method can be used to predict diverse properties of crystalline and amorphous materials, such as band gap energy and glass-forming ability.

## Introduction

Rational design of materials is the ultimate goal of modern materials science and engineering. As part of achieving that goal, there has been a large effort in the materials science community to compile extensive datasets of materials properties in order to provide scientists and engineers with ready access to the properties of known materials. Today, there are databases of crystal structures[1], superconducting critical temperatures,[2] physical properties of crystalline compounds,[3–6] and many other repositories containing useful materials data. Recently, it has been shown that these databases can also serve as resources for creating predictive models and design rules – the key tools of rational materials design.[7–13] These databases have grown large enough that the discovery of such design rules and models is impractical to accomplish by relying simply on human intuition and knowledge about material behavior. Rather than relying directly on intuition, machine learning offers the promise of being able to create accurate models quickly and automatically.

To date, materials scientists have used machine learning to build predictive models for a handful of applications.[14–28] For example, there are now models to predict the melting temperatures of binary inorganic compounds,[22] the formation enthalpy crystalline compounds,[15,16,29] which crystal structure is likely to form at a certain composition,[6,17,30–32] band gap energies of certain classes of crystals,[33,34] and the mechanical properties of metal alloys.[25,26] While these models demonstrate the promise of machine learning, they only cover a small fraction of the properties used in materials design and the datasets available for creating such models. For instance, no broadly-applicable, machine-learning-based models exist for the band gap energy or glass forming ability even though large-scale databases of these properties have existed for years.[3,35]

Provided the large differences between the approaches used in the literature, a systematic path forward to creating accurate machine learning models across a variety of new applications is not clear. While techniques in data analytics have advanced significantly, the development of routine methods for transforming raw materials data into the quantitative descriptions required for employing these algorithms has yet to emerge. In contrast, the chemoinformatics community benefits from a rich library of methods for describing molecular structures, which allow for standard approaches for deciding inputs into the models and, thereby, faster model development.[36–38] What is missing are similar flexible frameworks for building predictive models of material properties.

In this work, we present a general-purpose machine-learning-based framework for predicting the properties of materials based on their composition. In particular, we focus on the development of a set of attributes – which serve as input to the machine learning model – that could be reused for a broad variety of materials problems. Provided a flexible set of inputs, creating a new material property model can be reduced to finding a machine learning algorithm that achieves optimal performance – a well-studied problem in data science. Additionally, we employ a novel partitioning scheme to enhance the accuracy of our predictions by first partitioning data into similar groups of materials and training separate models for each group. We show that this method can be used regardless of whether the materials are amorphous or crystalline, the data is from computational or experimental studies, or the property takes continuous or discrete values. In particular, we demonstrate the versatility of our technique by using it for two distinct applications: predicting novel solar cell materials using a database of DFT-predicted properties of crystalline compounds and using experimental measurements of glass-forming ability to suggest new metallic glass alloys. Our vision is that this framework could be used as a basis for quickly creating models based on the data available in the materials databases and, thereby, initiate a major step forward in rational materials design.

## Results and Discussion

The results of this study are described in two major subsections. First, we will discuss the development of our method and the characterization of the attribute set using data from the OQMD. Next, we will demonstrate the application of this method to two distinct material problems.

### General Purpose Method to Create Materials Property Models

Machine learning (ML) models for materials properties are constructed from three parts: training data, a set of attributes that describe each material, and a machine learning algorithm to map attributes to properties. For the purposes of creating a general purpose method, we focused entirely on the attributes set because the method needs to be agnostic to the type of training data and because it is possible to utilize already-developed machine learning algorithms. Specifically, our objective is to develop a general set of attributes based on the composition that can be reused for a broad variety of problems.

The goal in designing a set of attributes is to create a quantitative representation that both uniquely defines each material in a dataset and relates to the essential physics and chemistry that influence the property of interest.[15,18] As an example, the volume of a crystalline compound is expected to relate to the volume of the constituent elements. By including the mean volume of the constituent elements as an attribute, a machine learning algorithm could recognize the correlation between this value and the compound volume, and use it to create a predictive model. However, the mean volume of the constituent elements neither uniquely defines a composition nor perfectly describes the volumes of crystalline materials.[39] Consequently, one must include additional attributes to create a suitable set for this problem. Potentially, one could include factors derived from the electronegativity of the compound to reflect the idea that bond distances are shorter in ionic compounds, or the variance in atomic radius to capture the effects of polydisperse packing. The power of machine learning is that it is not necessary to know which factors actually relate to the property and how before creating a model – those relationships are discovered automatically.

The materials informatics literature is full of successful examples of attribute sets for a variety of properties.[14–17,22,33,40] We observed that the majority of attribute sets were primarily based on statistics of the properties of constituent elements. As an example, Meredig, Agrawal *et al.* described a material based on the fraction of each element present and various intuitive factors, such as the maximum difference in electronegativity, when building models for the formation energy of ternary compounds.[16] Ghiringhelli *et al*. used combinations of elemental properties such as atomic number and ionization potential to study the differences in energy between zinc-blende and rocksalt phases.[15] We also noticed that the important attributes varied significantly depending on material property. The best attribute for describing the difference in energy between zinc-blende and rocksalt phases was found to be related to the pseudopotential radii, ionization potential, and electron affinity of the constituent elements.[15] In contrast, melting temperature was found to be related to atomic number, atomic mass, and differences between atomic radii.[22] From this we conclude that a general-purpose attribute set should contain the statistics of a wide variety of elemental properties in order to be adaptable.

Building on existing strategies, we created an expansive set of attributes that can be used for materials with any number of constituent elements. As we will demonstrate, this set is broad enough to capture a sufficiently-diverse range of physical/chemical properties in order to be used to create accurate models for many materials problems. In total, we use a set of 145 attributes, which are described in detail and compared against other attribute sets in the Supplementary Information, that fall into four distinct categories:

1. **Stoichiometric** attributes that depend only on the fractions of elements present and not what those elements actually are. These include the number of elements present in the compound and several $L^p$ norms of the fractions.
2. **Elemental Property Statistics**, which are defined as the mean, mean absolute deviation, range, minimum, maximum, and mode of 22 different elemental properties. This category includes attributes such as the maximum row on periodic table, average atomic number, and the range of atomic radii between all elements present in the material.
3. **Electronic Structure** attributes, which are the average fraction of electrons from the *s*, *p*, *d*, and *f* valence shells between all present elements. These are identical to the attributes used by Meredig, Agrawal *et al*.[16]
4. **Ionic Compound** attributes that include whether it is possible to form an ionic compound assuming all elements are present in a single oxidation state, and two adaptations of the fractional "ionic character" of a compound based on an electronegativity-based measure.[41]

For the third ingredient, the machine learning algorithm, we evaluate many possible methods for each individual problem. Previous studies have used machine learning algorithms including partial least-squares regression,[14,30] Least Absolute Shrinkage and Selection Operator (LASSO),[15,34,42] decision trees,[16,17] kernel ridge regression,[18–20,43] Gaussian process regression,[20–22,44] and neural networks.[23–25] Each method offers different advantages, such as speed or interpretability, which must be weighed carefully for a new application. We generally approach this problem by evaluating the performance of several algorithms to find one that has both reasonable computational requirements (i.e., can be run on available hardware in a few hours) and has low error rates in cross-validation – a process that is simplified by the availability of well-documented libraries of machine learning algorithms.[45,46] We often find that ensembles of decision trees (e.g., rotation forests[47]) perform best with our attribute set. These algorithms also have the advantage of being quick to train, but are not easily interpretable by humans. Consequently, they are less suited for understanding the underlying mechanism behind a material property but, owing to their high predictive accuracy, excellent choices for the design of new materials.

We also utilize a partitioning strategy that enables a significant increase in predictive accuracy for our ML models. By grouping the dataset into chemically-similar segments and training a separate model on each subset, we boost the accuracy of our predictions by reducing the breadth of physical effects that each machine learning algorithm needs to capture. For example, the physical effects underlying the stability intermetallic compounds are likely to be different than those for ceramics. In this case, one could partition the data into compounds that contain only metallic elements and another including those that do not. As we demonstrate in the examples below, partitioning the dataset can significantly increase the accuracy of predicted properties. Beyond using our knowledge about the physics behind a certain problem to select a partitioning strategy, we have also explored using an automated, unsupervised-learning-based strategy for determining distinct clusters of materials.[48] Currently, we simply determine the partitioning strategy for each property model by searching through a large number of possible strategies and selecting the one that minimizes the error rate in cross-validation tests.

### Justification for Large Attribute Set

The main goal of our technique is to accelerate the creation of machine learning models by reducing or eliminating the need to develop a set of attributes for a particular problem. Our approach was to create a large attribute set, with the idea that it would contain a diverse enough library of descriptive factors such it is likely to contain several that are well-suited for a new problem. To justify this approach, we evaluated changes in the performance of attributes for different properties and types of materials using data from the Open Quantum Materials Database (OQMD). As described in greater detail in the next section, the OQMD contains the DFT-predicted formation energy, band gap energy, and volume of hundreds of thousands of crystalline compounds. The diversity and scale of the data in the OQMD make it ideal for studying changes in attribute performance using a single, uniform dataset.

We found that the attributes which model a material property best can vary significantly depending on the property and type of materials in the dataset. To quantify the predictive ability of each attribute, we fit a quadratic polynomial using the attribute and measured the root mean squared error of the model. We found the attributes that best describe the formation energy of crystalline compounds are based on the electronegativity of the constituent elements (e.g., maximum and mode electronegativity). In contrast, the best-performing attributes for band gap energy are the fraction of electrons in the *p* shell and the mean row in the periodic table of the constituent elements. Additionally, the attributes that best describe the formation energy vary depending on the type of compounds. The formation energy of intermetallic compounds is best described by the variance in the melting temperature and number of *d* electrons between constituent elements, whereas compounds that contain at least one nonmetal are best modelled by the mean ionic character (a quantity based on electronegativity difference between constituent elements). Taken together, the changes in which attributes are the most important between these examples further supports the necessity of having a large variety of attributes available in a general-purpose attribute set.

It is worth noting that the 145 attributes described in this paper should not be considered the complete set for inorganic materials. The chemical informatics community has developed thousands of attributes for predicting the properties of molecules.[36] What we present here is a step towards creating such a rich library of attributes for inorganic materials. While we do show in the examples considered in this work that this set of attributes is sufficient to create accurate models for two distinct properties, we expect that further research in materials informatics will add to the set presented here and be used to create models with even greater accuracy. Furthermore, many materials cannot be described simply by average composition. In these cases, we propose that the attribute set presented here can be extended with representations designed to capture additional features such as structure (ex: Coulomb Matrix[18] for atomic-scale structure) or processing history. We envision that it will be possible to construct a library of general-purpose representations designed to capture structure and other characteristics of a material, which would drastically simplify the development of new machine learning models.

### Example Applications

In the following sections, we detail two distinct applications for our novel material property prediction technique in order to demonstrate its versatility: predicting three physically-distinct properties of crystalline compounds and identifying potential metallic glass alloys. In both cases, we use the same general framework, i.e., the same attributes and partitioning-based approach.

In each case, we only needed to identify the most-accurate machine learning algorithm and find an appropriate partitioning strategy. Through these examples, we discuss all aspects of creating machine-learning based models to design a new material: assembling a training set to train the models, selecting a suitable algorithm, evaluating model accuracy, and employing the model to predict new materials.

### Accurate Models for Properties of Crystalline Compounds

Density Functional Theory (DFT) is a ubiquitous tool for predicting the properties of crystalline compounds, but is fundamentally limited by the amount of computational time that DFT calculations require. In the past decade, DFT has been used to generate several databases containing the T = 0 K energies and electronic properties of ~$10^5$ crystalline compounds,[3–6,49] which each required millions of hours of CPU time to construct. While these databases are indisputably-useful tools, as evidenced by the many materials they have been used to design,[4,50–55] machine-learning-based methods offer the promise of predictions at several orders of magnitude faster rates. In this example, we explore the use of data from the DFT calculation databases as training data for machine learning models that can be used rapidly assess many more materials than what would be feasible to evaluate using DFT.

*Training Data:* We used data from the Open Quantum Materials Database (OQMD), which contains the properties of around 300,000 crystalline compounds as calculated using DFT.[3,4] We selected a subset of 228,676 compounds from OQMD that represent the lowest-energy compound at each unique composition to use as a training set. As a demonstration of the utility of our method, we developed models to predict the three physically-distinct properties currently available through the OQMD: band gap energy, specific volume, and formation energy.

*Method*: To select an appropriate machine learning algorithm for this example, we evaluated the predictive ability of several algorithms using 10-fold cross-validation. This technique randomly splits the dataset into 10 parts, and then trains a model on 9 partitions and attempts to predict the properties of the remaining set. This process is repeated using each of the 10 partitions as the test set, and the predictive ability of the model is assessed as the average performance of the model across all repetitions. As shown in Table 1, we found that creating an ensemble of reduced-error pruning decision trees using the random subspace technique had the lowest mean absolute error in cross-validation for these properties among the 10 ML algorithms we tested (of which, only 4 are listed for clarity).[56] Models produced using this machine learning algorithm had the lowest mean absolute error in cross validation, and had excellent correlation coefficients of above 0.91 between the measured and predicted values for all three properties.

As a simple test for how well our band gap model can be used for discovering new materials, we simulated a search for compounds with a band gap within a desired range. To evaluate our the ability of our method to locate compounds that have band gap energies within the target range, we devised a test where a model was trained on 90% of the dataset and then was tasked with selecting which 30 compounds in the remaining 10% were most likely to have a band gap energy in the desired range for solar cells: 0.9 – 1.7 eV.[57] For this test, we selected a subset of the OQMD that only includes compounds that have been reported to be possible to be made experimentally in the ICSD (a total of 25085 entries) so that only band gap energy, and not stability, needed to be considered.

For this test, we compared three selection strategies for finding compounds with desirable band gap energies: randomly selecting nonmetal-containing compounds (i.e., without machine learning), using a single model trained on the entire training set to guide selection, and a model created using the partitioning approach introduced in this manuscript. As shown in Figure 1, randomly selecting a nonmetal-containing compound would result in just over 12% of the 30 selected compounds to be within the desired range of band gap energies. Using a single model trained on the entire dataset, this figure dramatically improves to approximately 46% of selected compounds having the desired property. We found the predictive ability of our model can be increased to around 67% of predictions actually having the desired band gap energy by partitioning the dataset into groups of similar compounds before training. Out of the 20 partitioning strategies we tested, we found the best composite model works by first partitioning the dataset using a separate model trained to predict the expected range, but not the actual value, of the band gap energy (e.g., compounds predicted to have a band gap between 0 and 1.5 eV are grouped together), and then on whether a compound contains a halogen, chalcogen, or pnictogen. Complete details of the hierarchical model are available in the Supplementary

Information. By partitioning the data into smaller subsets, each of the individual machine learning models only evaluates compounds with similar chemistries (e.g. halogen-containing compounds with a band gap expected to be between 0 and 1.5 eV), which we found enhances the overall accuracy of our model.

Once we established the reliability of our model, we used it to search for new compounds (i.e., those not yet in the OQMD) with a band gap energy within the desired range for solar cells: 0.9 – 1.7 eV. To gain the greatest predictive accuracy, we trained our band gap model on the entire OQMD dataset. Then, we used this model to predict the band gap energy of compositions that were predicted by Meredig, Agrawal et al.[16] to be as-yet-undiscovered ternary compounds. Out of this list of 4500 predicted compounds, we found that 223 are likely to have favorable band gap energies. A subset with the best stability criterion (as reported in Ref. [16]) and band gap energy closest to 1.3 eV are shown in Table 3. As demonstrated in this example and by recent work from Sparks et al.,[58] having access to several machine learning models for different properties can make it possible to rapidly screen materials based on many design criteria. Provided the wide range of applicability of the machine learning technique demonstrated in this work and the growing availability of material property data, it may soon be possible to screen for materials based on even more properties than those considered here using models constructed based on several different datasets.

### Locating Novel Metallic Glass Alloys

Metallic glasses possess a wide range of unique properties, such as high wear resistance and soft magnetic behavior, but are only possible to create at special compositions that are difficult to determine *a priori*.[59] The metallic glass community commonly relies on empirical rules (e.g., systems that contain many elements of different sizes are more likely to form glasses[60]) and extensive experimentation in order to locate these special compositions.[56] While searches based on empirical rules have certainly been successful (as evidenced by the large variety of known alloys[61]), this conventional method is known to be slow and resource-intensive.[62] Here, we show how machine learning could be used to accelerate the discovery of new alloys by using known experimental datasets to construct predictive models of glass forming ability.

*Data:* We used experimental measurements taken from "Nonequilibrium Phase Diagrams of Ternary Amorphous Alloys," a volume of the Landolt-Börnstein collection.[33] This dataset contains measurements of whether it is possible to form a glass using a variety of experimental techniques at thousands of compositions from hundreds of ternary phase diagrams. For our purposes, we selected 5369 unique compositions where the ability to form an amorphous ribbon was assessed using melt spinning. In the event that multiple measurements for glass forming ability were taken at a single composition, we assume that it is possible to form a metallic glass if at least one measurement found it was possible to form a completely-amorphous sample. After the described screening steps, 70.8% of the entries in the training dataset correspond to metallic glasses.

*Method:* We used the same set of 145 attributes as in the band gap example and ensembles of Random Forest classifiers[63] created using the random subspace technique as the machine learning algorithm, which we found to be the most accurate algorithm for this problem. This model classifies the data into two categories (i.e., can and cannot form a metallic glass) and computes the relative likelihood that a new entry would be part of each category. For the purposes of validating the model, we assume any composition predicted to have a greater than 50% probability of glass formation to be a positive prediction of glass forming ability. Using a single model trained on the entire dataset, we were able to create a model with 90% accuracy in 10-fold cross-validation.

As a test of the ability of our method to predict new alloys, we removed all entries that contained exclusively Al, Ni, and Zr (i.e., all Al-Ni-Zr ternary compounds, and any binary formed by any two of those elements) from our training dataset and then predicted the probability of an alloy being able to be formed into the amorphous state for the Al-Ni-Zr ternary system. As shown in Figure 2a, it is possible to form amorphous ribbons with melt spinning in one region along the Ni-Zr binary and in a second, Al-rich ternary region. Our model is able to accurately predict both the existence of these regions and their relative locations (see Figure 2b), which shows that models created using our method could serve to accurately locate favorable compositions in yet-unassessed alloy systems.

We further validated the ability of our models to extrapolate to alloy systems not included in the training set by iteratively using each binary system as a test set. This procedure works by excluding all alloys that contain both of the elements in the binary, training a model on the remaining entries, and then predicting the glass-forming ability of the alloys that were removed. For example, if the Al-Ni binary were being used as a test set, then $Al_{50}Ni_{50}$ and $Al_{50}Ni_{25}Fe_{25}$ would be removed but $Al_{50}Fe_{50}$ and $Al_{50}Fe_{25}Zr_{25}$ would not. This process is then repeated for all 380 unique binaries in the dataset. We measured that our model has an 80.2% classification accuracy over 15318 test entries where 71% of entries were measured to be glasses – in contrast to the 90.1% measured in 10-fold cross-validation with a similar fraction of glasses in the test set. We also found that by training separate models for alloys that contain only metallic elements and those that contain a nonmetal/metalloid it is possible to slightly increase the prediction accuracy to 80.7% - a much smaller gain than that observed in the band gap example (23%). Overall, this exclusion test strongly establishes that our model is able to predict the glass forming ability in alloy systems that are completely unassessed.

In order to search for new candidate metallic glasses, we used our model to predict the probability of glass formation for all possible ternary alloys created at 2 at% spacing by any combination of elements found in the training set. Considering that the dataset included 51 elements, this space includes approximately 24 million candidate alloys, which required approximately 6 hours to evaluate on 8, 2.2 GHz processors. In order to remove known alloys from our prediction results, we first removed all entries where the $L_1$ distance between the composition vector (i.e., $\langle x_H, x_{He}, x_{Li}, ...\rangle$) of the alloy and any amorphous alloy in the training set was less than 30 at%. We then found the alloys with the highest predicted probability of glass formation in each binary and ternary. Eight alloys with the highest probability of glass formation are shown in Table 3. One top candidate, $Zr_{0.38}Co_{0.24}Cu_{0.38}$, is particularly promising considering the existence of Zr-lean Zr-Co and Zr-Cu binary alloys and Zr-Al-Co-Cu bulk metallic glasses.[64] To make the ability to find new metallic glasses openly available to the materials science community, we have included all of the software and data necessary to use this model in the Supplementary Information and created an interactive, web-based tool.[65]

## Conclusions

In this work, we introduced a general-purpose machine learning framework for predicting the properties of a wide variety of materials and demonstrated its broad applicability via illustration of two distinct materials problems: discovering new potential crystalline compounds for photovoltaic applications and identifying candidate metallic glass alloys. Our method works by using machine learning to generate models that predict the properties of a material as a function of a wide variety of attributes designed to approximate chemical effects. The accuracy of our models is further enhanced by partitioning the dataset into groups of similar materials. In this manuscript, we show that this technique is capable of creating accurate models for properties as different as the electronic properties of crystalline compounds and glass formability of metallic alloys. Creating new models with our strategy requires only finding which machine learning algorithm maximizes accuracy and testing different partitioning strategies, which are processes that could be eventually automated.[66] We envision that the versatility of this method will make it useful for a large range of problems, and help enable the quicker deployment and wider-scale use machine learning in the design of new materials.

## Methods

All machine learning models were created using the Weka machine learning library.[45] The Materials Agnostic Platform for Informatics and Exploration (Magpie) was used to compute attributes, perform the validation experiments, and run searches for new materials. Both Weka and Magpie are available under open-source licenses. The software, training datasets, and input files used in this work are provided in the Supplementary Information associated with this manuscript.

## Acknowledgments

All authors gratefully acknowledge primary funding support from DOC NIST award 70NANB14H012 (CHiMaD). Additionally, AA and AC were supported in part by the following grants: DARPA SIMPLEX award N66001-15-C-4036; NSF awards IIS-1343639 and CCF-1409601; DOE award DESC0007456; and AFOSR award FA9550-12-1-0458. LW was partially supported by

# Contributions

CW conceived the project, and jointly developed the method with LW, AA and AC. LW wrote all software and performed the necessary calculations with help and guidance from AA and AC. LW led the manuscript writing, with all other authors contributing.

# Competing Interests

The authors declare no conflict of interest.

# Figures

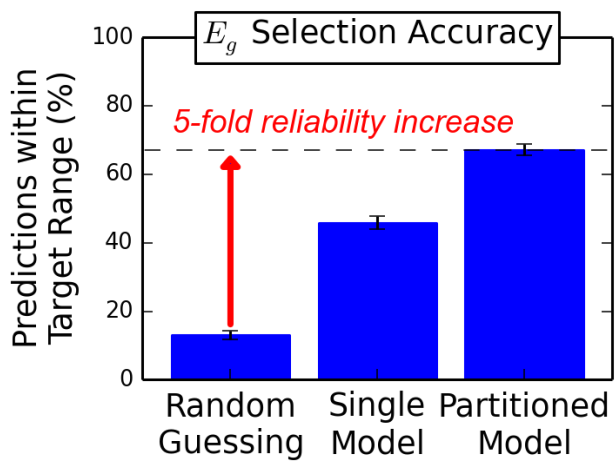

**Figure 1.** Performance of three different strategies to locate compounds with a band gap energy within a desired range: randomly-selecting nonmetal-containing compounds, and two strategies using the machine-learning-based method presented in this work. The first machine learning strategy used a single model trained on the computed band gap energies of 22667 compounds from the ICSD. The second method a model created by first partitioning the data into groups of similar materials, and training a separate model on each subset. The number of materials that were actually found to have a band gap within the desired range after 30 guesses was over 5 times larger when using our machine learning approach than when randomly selecting compounds. Error bars represent the 95% confidence interval.

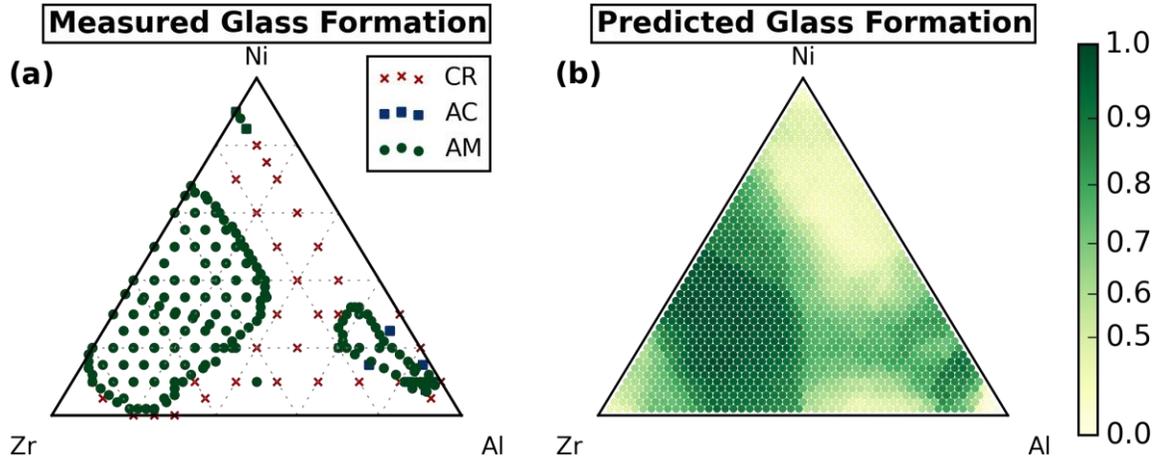

**Figure 2.** (a) Experimental measurements of metallic glass forming ability in the Al-Ni-Zr ternary, as reported in Ref. [35]. Green circles (AM) mark compositions at which it is possible to create a fully-amorphous ribbon via melt spinning, blue squares (AC) mark compositions at which only a partially-amorphous ribbon can be formed, and red squares (CR) mark compositions where it is not possible to form any appreciable amount of amorphous phase. (b) Predicted glass forming ability from our machine learning model. Points are colored based on relative likelihood of glass formation, where 1 is the mostly likely and 0 is the least. The model used to make these predictions was developed using the methods outlined in this work, and was not trained on any measurements from the Al-Ni-Zr ternary or any of its constituent binaries.

# Tables

**Table 1.** Comparison of the ability of several machine learning algorithms to predict properties of materials from the OQMD. Data represents the mean absolute error in a 10-fold cross-validation test of a single model trained on the properties predicted using DFT of 228,676 crystalline compounds.

|  | | Machine Learning Algorithm | | | |
|---|---|---|---|---|---|
|  | | Linear Regression | Reduced-Error Pruning Tree (REPTree) | Rotation Forest[47] + REPTree | Random Subspace[56] + REPTree |
| **Property** | Volume ($Å^3$/atom) | 1.22 | 0.816 | 0.593 | 0.563 |
|  | Formation Energy (eV/atom) | 0.259 | 0.126 | 0.0973 | 0.0882 |
|  | Band gap Energy (eV) | 0.202 | 0.0701 | 0.0643 | 0.0645 |

**Table 2.** Compositions and predicted band gap energies of materials predicted using machine learning to be candidates for solar cell applications. Compositions represent the nominal compositions of novel ternary compounds predicted by using methods developed in Ref. [16]. Band gap energies were predicted using a machine learning model trained on DFT band gap energies from the OQMD[3] using methods described in this work.

| Composition | $E_g$ (eV) |
|---|---|
| $ScHg_4Cl_7$ | 1.26 |
| $V_2Hg_3Cl_7$ | 1.16 |
| $Mn_6CCl_8$ | 1.28 |
| $Hf_4S_{11}Cl_2$ | 1.11 |
| $VCu_5Cl_9$ | 1.19 |

**Table 3.** Compositions of candidate metallic glass alloys predicted using a machine learning model trained on experimental measurements of glass forming ability. These alloys were predicted to have the highest probability being able to be formed into an amorphous ribbon via melting spinning out of 24 million candidates.

| Alloy Composition | |
|---|---|
| $Zr_{0.38}Co_{0.24}Cu_{0.38}$ | $Hf_{0.7}Si_{0.16}Ni_{0.14}$ |
| $V_{0.16}Ni_{0.64}B_{0.2}$ | $Hf_{0.48}Zr_{0.16}Ni_{0.36}$ |
| $Zr_{0.46}Cr_{0.36}Ni_{0.18}$ | $Zr_{0.48}Fe_{0.46}Ni_{0.06}$ |
| $Zr_{0.5}Fe_{0.38}W_{0.12}$ | $Sm_{0.22}Fe_{0.54}B_{0.24}$ |